\documentclass[]{interact}

\usepackage{epstopdf}
\usepackage{subcaption} 
\usepackage{algorithm}
\usepackage{algpseudocode}
\usepackage{eurosym}

\usepackage{multirow}

\usepackage[hidelinks]{hyperref}
\usepackage[natbibapa,nodoi]{apacite}

\theoremstyle{plain}

\theoremstyle{definition}

\theoremstyle{remark}

\begin{document}
\immediate\write18{cp `kpsewhich apacite.bst` .}

\title{On the Transition to an Auction-based Intelligent Parking Assignment System}


\author{
\name{Levente Alekszejenkó\textsuperscript{a}\thanks{CONTACT Levente Alekszejenkó. Email: alekszejenko.levente@vik.bme.hu} and Tadeusz Dobrowiecki\textsuperscript{a}}
\affil{\textsuperscript{a}Department of Artificial Intelligence and Systems Engineering,\\ Budapest University of Technology and Economics,\\ Műegyetem rkp. 3., H-1111 Budapest, Hungary}
}

\maketitle

\begin{abstract}
Finding a free parking space in a city has become a challenging task over the past decades. A recently proposed auction-based parking assignment can alleviate cruising for parking and also set a market-driven, demand-responsive parking price. However, the wide acceptance of such a system is far from certain.

To evaluate the merits of auction-based parking assignment, we assume that drivers have access to a smartphone-based reservation system prior to its mandatory introduction and thus have the opportunity to test and experience its merits voluntarily. We set our experiment as Eclipse SUMO simulations with different rates of participants and non-participants to check how different market penetration levels affect
the traffic flow, the performance of the auction-based assignment system, and the financial outcomes. The results show that the auction-based system improves traffic flow with increasing penetration rates, allowing participants to park gradually closer to their preferred parking lots. However, it comes with a price; the system also increases parking expenditures for participants. Interestingly, non-participating drivers will face even higher parking prices. Consequently, they will be motivated to use the new system.
\end{abstract}


\section{Introduction}
In recent decades, finding a vacant curbside parking space in a city's central business district (CBD) has become a serious problem. According to \cite{shoup}, 15-68\% of the vehicles might cruise to find a free parking space, seriously increasing the harmful emissions of transportation and requiring 3.3-15.4 minutes additional travel time. Consequently, to address this problem, many cities invested millions of dollars in intelligent parking systems, as \cite{lin} summarized. Considering that most cities cannot afford to implement such complex solutions, using multi-agent artificial intelligence methods can be an affordable alternative. For example, \cite{chen} introduce a Vickrey-Clarke-Groves (VCG) auction-based intelligent parking assignment that, in addition to providing a parking assignment, can also optimize parking prices to respond to current demand. \cite{shoup} states, this market-driven price would be optimal for curbside parking.

\begin{figure}
    \centering
    \subfloat[]{
    \resizebox*{10cm}{!}{\includegraphics{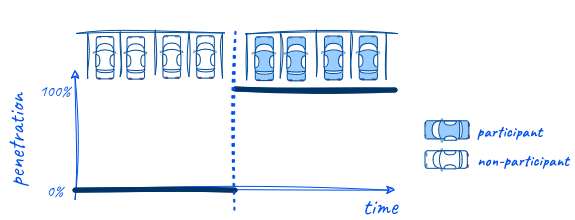}}~\label{fig:mandatory_system}} \\
    \subfloat[]{
    \resizebox*{6cm}{!}{\includegraphics{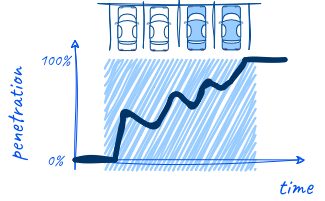}}~\label{fig:transition_parking}}
    
    \caption{(a) Parking lot operators might require drivers to mandatory use the  intelligent parking lot assignment system. In this case, penetration rate rises from 0\% to 100\% instantly. (b) If drivers can voluntarily participate in a novel system, it leads to a transition with various penetration levels.} \label{fig:transition}
\end{figure}

Unfortunately, drivers are not necessarily willing to pay an increased parking price, as the questionnaires of \cite{hashimoto, vidovic, simicevic} point out. Therefore, it is unlikely that the city's governance would require venture to an auction-based parking assignment system as in Figure~\ref{fig:mandatory_system}. However, if users voluntarily perceive the usefulness of an auction-based parking assignment, they are likely to accept and use it, according to the technology acceptance survey of \cite{marangunic}. A driver would find such a system beneficial if it saves time, prevents cruising, or, perhaps, saves some money. If the system can prove its merits, it will attract more and more users. In contrast to a mandatory system, this process will create a transition between a conventional and fully automated parking assignment system, see Figure~\ref{fig:transition_parking}. However, many studies, for example, \cite{yu, ozioko}, investigate the impacts of mixed traffic of human drivers and autonomous vehicles, but only a handful of articles focus on intelligent parking solutions with participating and not participating drivers.

We aim to fill this research gap by investigating the transition from the traditional parking system to an auction-based parking assignment solution. With different ratios of participating and non-participating drivers (penetration rates), we run microscopic traffic simulations to answer the following questions:
\begin{itemize}
	\item How does the market penetration of an auction-based parking assignment influence the traffic of a CBD?
	\item Will an auction-based parking assignment be beneficial enough to attract more users right after its market introduction?
	\item How does the market penetration of an auction-based parking assignment influence parking costs?
\end{itemize}

The rest of the paper is organized as follows. In section~\ref{sec:literature}, we address related works. In section~\ref{sec:auction}, we propose using an ascending bid auction for parking lot assignment. Section~\ref{sec:simulation} gives some insight into the simulation scenario used for the evaluation. In Section~\ref{sec:results}, we answer the above research questions using the results of the simulations. 
Finally, section~\ref{sec:conclusion} concludes this paper.

\section{Related Works}
\label{sec:literature}
Pricing curbside parking has required the precise planning of economists, engineers, and politicians since installing the first parking meter in 1935. According to \cite{inci}, if curbside parking is underpriced, it will lead to an overdemand, forcing drivers to look for a vacant parking space. On the other hand, with the market-driven, temporal-spatial parking fees, municipalities can eliminate the cruising for parking.

\cite{waraich} presents a type of genetic algorithm for the temporal-spatial optimization of parking prices, while \cite{anderson} presents a numerical solution. For online optimization, \cite{chen} proposes a Vickrey-Clarke-Groves (VCG) auction mechanism deployed as a smartphone application. In addition to smartphone applications, auction-based parking assignment systems can also utilize the capabilities of vehicular fog computing as in the works of \cite{zhang2018, zhang2019}. As VCG does not provide a budget balance, \cite{cheng} introduced scale control to satisfy this property by constraining the number of agents who can be assigned to a parking space. For the same reason, \cite{xiao2018} proposed a demander competition padding method in their double-auction mechanism. Despite the usual approach, in which drivers compete for parking lots, \cite{tan} describes a reverse Vickrey auction, in which parking spaces compete for vehicles.

Although intelligent parking solutions can eliminate cruising for parking and improve the traffic flow, as the survey of \cite{khalid} points out, data privacy is still an open question. As \cite{shoup} summarizes, the optimal parking spot minimizes the total time and monetary costs of parking and walking. Thus, reporting bids for parking spaces in a VCG auction can reveal drivers' private information about their financial and health status, e.g., their willingness or ability to walk. Building on the work of \cite{demange}, \cite{bansal} present that decentralized simultaneous independent ascending bid auctions (SIA) with local greedy bidding (LGB) allow the integration of independent online auctions running at the same time. In this approach, the decentralized parking lot auctioneer agents will only obtain a couple of bits of information about whether a driver finds this optimal in a given time for a given bid or not. \cite{rizvi} implement SIA for intelligent parking assignment and test it using a simulation framework. In this paper, we follow a similar approach. We implement a variant of a monetary English SIA with LGB, see section~\ref{sec:auction}, and evaluate its capabilities with different penetration rates by simulations in Eclipse SUMO, which simulation tool was developed by \cite{sumo}.

However, as the work of \cite{khalid} reflects, numerous papers focus on intelligent parking solutions; only a few articles analyze their effect on society. \cite{chen2024} concern that auction-based parking assignment might lead to inequity in the market. Higher-income drivers who can and are willing to pay more for parking could prevent lower-income drivers from finding a suitable parking space. As every novel solution, a voluntary auction-based parking lot assignment system requires some time for technological acceptance. In this study, we call the ratio of participating drivers to the total number of drivers the \emph{penetration rate}. As an intelligent parking solution might provide different results in an early market stage, when the penetration rate is low, the researchers shall also evaluate whether innovators and early adapters would enjoy some benefits of the proposed system. Moreover, according to \cite{meade} the operators of the system should also plan the product and marketing strategy that reflect the needs of the stakeholders in each phase of the product's life cycle. \cite{chai} evaluate the penetration rate of a dynamic rerouting system to minimize travel and parking costs in parking garages. They found that higher penetration rates lower parking-related costs and improve travel times. \cite{ni} propose a parking reservation system and check its performance with different penetration rates. According to their results, regular vehicles have to travel and their drivers have to walk more as the penetration level of the proposed system increases. Even a less complex guidance system, which only informs drivers at intersections whether they can find a vacant space on a downstream link, significantly reduces the average search distance for the early market, while laggards will suffer a significant increase at higher penetration rates according to \cite{leclercq}.

\section{The Auction Mechanism}
\label{sec:auction}

\begin{figure}
    \centering
    \resizebox*{13cm}{!}{\includegraphics{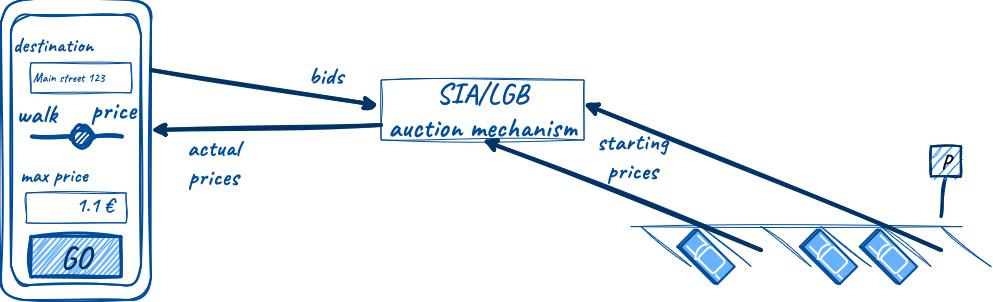}}
    \caption{Overview of the assumed auction-based parking assignment system. Drivers can set their destination, preference of walking distance over parking price, and the maximum parking price they are willing to pay in a smartphone app. This application connects to the simultaneous independent ascending auction service to report their local greedy bids (SIA/LGB). An auction corresponds to each free parking space in the area and has a predefined starting price.} \label{fig:use_case}
\end{figure}

Similarly to \cite{chen}, we assume that the auction-based parking system can be deployed as a smartphone application as smart phones are available gadgets for the vast majority of people; see the statistics of \cite{sp_statistics}. Drivers can decide to use this mobile app to guide them to a parking facility that has a vacant parking space assigned to them, see Figure~\ref{fig:use_case}. Drivers would be able to give their destinations, their preference between longer walking distances and higher parking fees, and the maximum amount they are willing to pay for parking. The application combines the functions of a navigation and a parking reservation service, and guides the human drivers to a vacant parking space. The application in the background automatically bids on SIA auctions according to the given preference of the driver.

SIA auctions run independently as in \cite{bansal}; hence, each parking operator, which currently has some free spaces, can organize its auction $A_i$ with a predefined minimum starting price $p_i^{(0)}$. The auctioneer service schedules the auctions, i.e., in the adopted implementation\footnote{For details of the implementation, please visit our GitHub repository:\\ 
\url{https://github.com/alelevente/penparking}
}, auctions start with a 15~s period time, and as their execution duration is bounded by the product of the number of free parking lots and the biggest difference between the starting prices and highest parking price that the drivers are willing to pay, we assume that the auctions can be calculated within these 15~s periods.

\begin{figure}
    \centering
    \resizebox*{14cm}{!}{\includegraphics{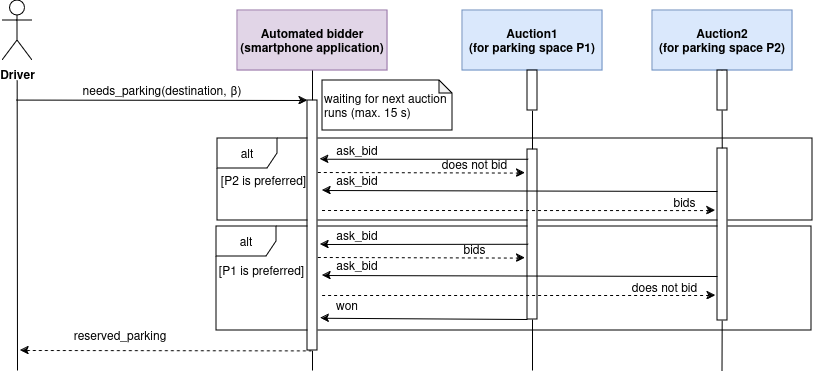}}
    \caption{An example of the SIA auction running. Drivers request parking spaces via a smartphone application. The application bids on simultaneous auctions to reserve a parking lot based on the preference ($\beta$) and the destination of the drivers. When the auctions terminates, drivers will be informed where they can find their reserved parking space.} \label{fig:seqd}
\end{figure}

Participants' smartphones register at these auctions to bid for parking lots. After that, in rounds, each auction $A_i$ asks each registered smartphone agent $j$ whether it bids for parking at the current $p_i$ price, see Fig.~\ref{fig:seqd}. If a participant bids, auction $A_i$ increases its current price by a small amount $\epsilon$, $p_i' \gets p_i + \epsilon$. If no one bids for a while, the auction terminates. The winner will be able to occupy the parking space of $A_i$. This method corresponds to online auctions on the Internet. In this paper, for the sake of simplicity, we treat winning bids as the actual parking prices, regardless of the duration of parking.

Smartphone agents use the local greedy bidding (LGB) strategy of \cite{bansal} to obtain \emph{at most one} free parking space. Consequently, an agent is allowed to submit the current bid in no more than one auction from the ones running simultaneously. Following \cite{shoup}, we consider two factors when an agent $j$ decides if it will bid or not. The $d_{i,j}$ measures the \emph{driving} distance between the parking lot offered at auction $A_i$ and the destination of participant $j$. As one can use shortcuts and walk through passages on foot, driving distance is always an overestimate of the walking distance between two points in a city. Another important factor in parking lot selection is the $p_i$ parking price. For certain activities (e.g., when shopping or when a deliveryman brings us a package), drivers prefer the closest possible parking options; while for other activities (e.g., when participating in some leisure activities) we might prefer cheaper (but probably more distant) parking lots. To express this attitude towards parking lots, we introduce a $\beta \in (0,1]$ \emph{attitude factor} to weigh between parking prices and walking distances.

For technical reasons, let $p_{max}$ denote the maximum current price in the auctions and $d_{j, max}$ the driving distance from the most remote parking lot to the destination of $j$. Then, we calculate the total $c_{i,j}$ cost of parking for participant $j$ at the parking space being sold by auction $A_i$, considering attitude factor $\beta$ as:
\begin{equation}
	c_{i,j} = \beta \frac{p_{i}}{p_{max}} + (1-\beta)\frac{d_{i,j}}{d_{j, max}}. \label{eq:costs}
\end{equation}

By setting a higher $\beta$, drivers can express their attitude toward cheaper, yet possibly more distant parking prices; and vice-versa, with a lower $\beta$, drivers can favor closer parking lots.

Each agent $j$ tries to minimize the $c_{i,j}$ total parking cost (consisting of monetary parking expenditures and walking distances); hence, the utility function of agent $j$ bidding on auction $i$ can be expressed as $U_{i,j} = 1-c_{i,j}$. This yields the $\Pi_j$ preferred auction of agent $j$ as:
\begin{equation}
	\Pi_j = \arg \min_i c_{i,j}. \label{eq:preference}
\end{equation}

Therefore, if agent $j$ gets overbid, it shall bid next at the auction $\Pi_j$. Necessarily (otherwise the auctions would not terminate), everyone has some $p_{max}$ limit (personal valuation) they can spend on parking. In our study, we used $p_{max} = 5.0$~\euro~parking valuation and an $\epsilon = 0.05$~\euro~bid step. 
The auction results provide an assignment between the participating vehicles and the parking lots and also set a market-driven pricing.

If a participating driver $j$ (represented by a smartphone agent) wins an auction, the smartphone agent can plan a route to the won and virtually reserved parking space\footnote{Technically, during simulations, we cancel the reservation if a participating vehicle leaves the parking space assigned to it, or if the reservation was physically unsuccessful. The latter happens, if a non-participant occupies the reserved space before the participant driver would have reached it.}.

On the other hand, non-participant drivers find vacant parking spaces by the traditional method; they cruise for parking and pay the predefined parking prices. Moreover, as they do not use the assignment system, they cannot be aware that they might occupy a space reserved for a participant's vehicle. In this case, we assume that the participating driver returns to the traditional parking-seeking method, cruises until finding a free space, and pays the original fee.

\section{Simulations}
\label{sec:simulation}

To analyze how different penetration rates affect an SIA auction-based parking assignment system, we conducted various simulations. To model and control the simulated vehicles as individual intelligent agents, we used a microscopic traffic simulation tool, called Eclipse SUMO, developed by \cite{sumo}. The Eclipse SUMO simulates the vehicles on an agent level, calculating their longitudinal and lateral movements, and considering their interaction with each other and the infrastructure, such as traffic lights or parking lots.

\subsection{Parking Simulation in Eclipse SUMO}
\label{sec:cruising_sim}

The Eclipse SUMO provides simulation of parking, see \cite{sumoparking}. Firstly, one shall define \texttt{ParkingAreas}\footnote{See \href{https://sumo.dlr.de/docs/Simulation/ParkingArea.html}{ParkingArea documentation of Eclipse SUMO}, accessed: 2025. June 11.} in the simulated road network. Then, by assigning \texttt{
stops}\footnote{See \href{https://sumo.dlr.de/docs/Definition\_of_Vehicles,\_Vehicle_Types,\_and\_Routes.html\#stops\_and\_waypoints}{vehicle definition documentation of Eclipse SUMO}, accessed: 2025. June 11.} in \texttt{ParkingAreas} to the simulated vehicles, we can instruct them to occupy a free parking space and spend there some predefined time.

\begin{figure}
    \centering
    \subfloat[]{
    \resizebox*{5cm}{!}{\includegraphics{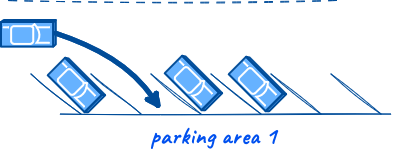}\label{fig:freeparking}}} \hspace{10pt}
    \subfloat[]{
    \resizebox*{8cm}{!}{\includegraphics{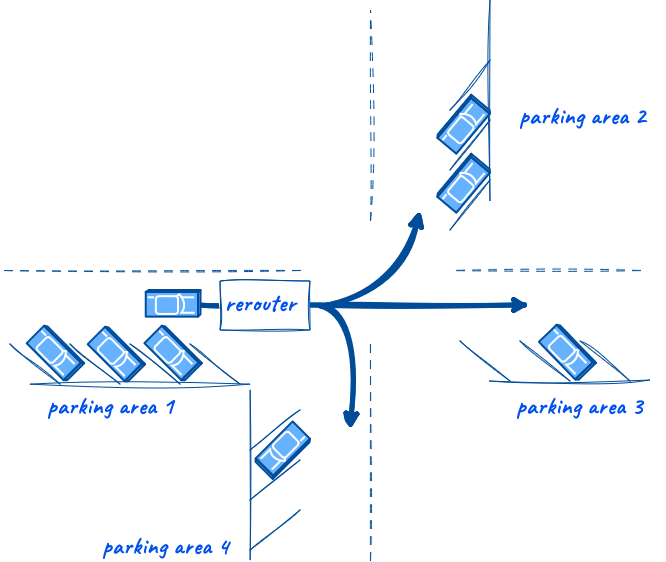}}\label{fig:rerouter}}
    
    \caption{Parking simulation in Eclipse SUMO. (a) if there are free spaces in a vehicle's designated parking area, the vehicle occupies the first free space. (b) when its designated parking area is full, a vehicle can be rerouted via a \texttt{ParkingAreaRerouter} to the neighboring parking areas.} \label{fig:parkingsim}
\end{figure}

During simulations, as depicted in Figure~\ref{fig:parkingsim}, vehicles try to occupy the first free parking space in the parking area defined by a \texttt{stop} event. However, their designated parking lot might be full. In such cases, by placing \texttt{ParkingAreaRerouters}\footnote{See \href{https://sumo.dlr.de/docs/Simulation/Rerouter.html}{ParkingAreaRerouter documentation of Eclipse SUMO}, accessed 2025. June 11.}, the Eclipse SUMO can dispatch the vehicles towards (by default, with equal probability) the nearest parking areas (inside two blocks' distance) that currently offer at least one free parking space. Consequently, Eclipse SUMO simulates the queue- and block-face-based model proposed by \cite{queuing} to simulate \emph{cruising for parking}.

\subsection{Parking Scenario}
\label{sec:parkscen}
To draw a general conclusion, we created an abstract simulation scenario rather than selecting one of the few real-world ones. In this way, we can avoid network-specific biases. Through optimization, we establish a scenario in which the number of available parking lots and parking demand are roughly in balance. Consequently, during a simulation run, it leads to a stable \emph{parking occupancy}, the ratio of occupied parking lots to the total number of parking lots. If we increased the traffic, the whole simulation would run into a deadlock due to the vehicles that cannot find a free parking space. On the other hand, if we added more parking lots, vehicles would not cruise for parking.

Specifically, the road network in our simulation scenario is a $6 \times 6$ grid network, with junctions spaced 100~m apart, corresponding to a typical block size. We use this model to simulate the traffic of approximately 4~hours of a central business district (CBD) that offers on-street parking for short-term visits (uniformly random between 15~and 45~minutes), e.g., going to a store, having a meal in a restaurant, or meeting in an office. The innermost part of the CBD attracts more of the 11520~simulated visitors than the outskirts of the simulated network. 

In our simulation, we model the road network as a directed graph, representing the junctions as graph nodes and directed streets as graph \emph{edges}. The outermost edges of the road network attract the visitors with a probability of $\frac{1}{336}$, while the innermost edges with $\frac{5}{336}$. The attraction probability increases evenly by $\frac{1}{336}$ with each junction while getting closer to the nucleus of the CBD.

To model that a vehicle probably originated outside of the CBD, we use a different approach to define the distribution of vehicle origins. The outmost edges are 9-times more likely to be the origin of the vehicles compared to the innermost edges. The other edges' probabilities are proportional with their Euclidean distance from the geometric center of the road network. Therefore, vehicles are likely to appear on the outer edges of the road network, move toward the center of the CBD, and stop to park there for a while. After that, they return to their street of origin and leave the simulation, see in Figure~\ref{fig:grid_distance}.

On each edge, we defined \texttt{ParkingAreas} in Eclipse SUMO. All of these areas offer 15~parking spaces parallel to the edge itself, with each parking space measuring 6.6~m in length.\footnote{Modeling accidents in parking lots and considering maneuvering times are currently out of scope of the research; hence, parking spaces' length is only a visual property.}

\begin{figure}[tb]
    \centering
    \resizebox*{12.5cm}{!}{\includegraphics{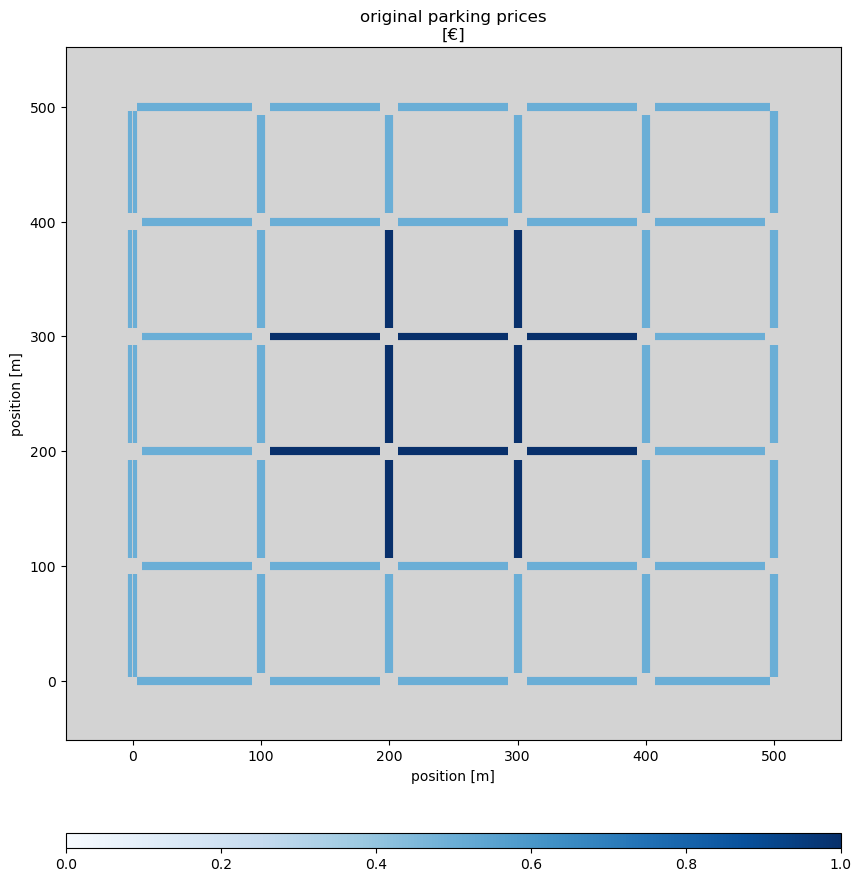}}
    \caption{Road network and parking zones in the simulated scenario. Colors correspond to the original parking prices in the two parking zones.} \label{fig:map}
\end{figure}

We assume that the municipality or parking operators have already established a parking pricing scheme. In this scheme, there are two parking zones; the inside parking spaces are more expensive (having a 1\euro~parking price), while the outer parking lots are cheaper (with 0.5\euro), as shown in Figure~\ref{fig:map}.

To model that there is a part of drivers who do not or cannot care about parking fees, e.g., couriers, tradespeople, the elderly, or disabled people, we defined three \emph{measurement cases} with different mixtures of $\beta$ settings in the population, see Table~\ref{tab:mix_definitions}. For example, in the \texttt{MIX10} case, 10\% of the drivers have a small $\beta=0.01$ setting, preferring the closer parking lots over the more distant ones; while the rest of the drivers have a more neutral $\beta=0.5$ preference. With the \texttt{MIX10} case, we can model a population with high parking price aversion or the behavior on a Summer day with fair weather, when people are more likely to opt for cheaper parking prices and will walk longer distances. On the other hand, the \texttt{MIX50} case represents a population that prefers more comfortable, but probably more expensive parking options, or a natural human behavior in unfavorable weather conditions. The \texttt{MIX25} case is between the former two, in which a significant part of drivers does not care about parking costs. However, they are far from being a majority of the population.

\begin{table}[tb]
    \tbl{Population definition in various measurement cases.}
    {\begin{tabular}{l|cc} \toprule
         measurement case & $\mathbb{P}(\beta = 0.01)$ & $\mathbb{P}(\beta = 0.5)$\\ \midrule
         \texttt{MIX10} & 0.1 & 0.9 \\
         \texttt{MIX25} & 0.25 & 0.75 \\
         \texttt{MIX50} & 0.5 & 0.5 \\
         \bottomrule
    \end{tabular}}
    \label{tab:mix_definitions}
\end{table}

\subsection{Parking Behaviors}
To evaluate the capabilities of the proposed auction-based parking assignment system, we experimented with three levels of parking guidance systems, as shown in Table~\ref{tab:reservation_system}: a traditional, \emph{baseline} system that does not provide parking lot occupancy information, or opportunity for a parking reservation. An \emph{information} system which provides real-time parking lot occupancy data, but it still does not offer parking reservation. Furthermore, we implemented and tested the proposed \emph{auction}-based parking reservation system. Together, we set up $3 \times (1+2\times4)$ simulation cases, to test with each measurement case with each systems. As the information system and the auction-based parking assignment system is assumed to require using a smartphone application, we tested them on four levels of market penetration.

\begin{table}[tb]
    \tbl{Parking behaviors.}
    {\begin{tabular}{l|cc} \toprule
         parking system & real-time parking occupancy information & parking reservation\\ \midrule
         baseline & --- & --- \\
         information & $\checkmark$ & --- \\
         auction & $\checkmark$ & $\checkmark$ \\
         \bottomrule
    \end{tabular}}
    \label{tab:reservation_system}
\end{table}

By comparing the auction to the information system, we can observe the merits of the assignment system on each penetration levels. By comparing the information system, to the baseline system, we can observe how a simple parking occupancy information helps drivers to find vacant parking spaces in or near their preferred parking lot. By comparing the auction system to the baseline system, one can evaluate the merits of a parking assignment system which also incorporates a dynamic parking pricing scheme.

\subsubsection{Baseline}
With the classical, \emph{baseline} behavior, the simulated vehicles drive through their route between their sampled origin and their vehicular destination according to their preferred parking lot in the actual measurement case. However, the vehicles have no information about or a reservation for parking lots. Consequently, there is no guarantee that they will find an empty parking space in their preferred parking lot. If they cannot park there, vehicles will start cruising as described in section~\ref{sec:cruising_sim}.

\subsubsection{Information}
\label{sec:info_behav}
We also ran simulations to compare the merits of the auction-based reservation to a simple parking guidance system. To this end, drivers using the \emph{information} system check parking lot occupancies right before departure. By using such a system, they can iterate over their parking preference list, ordered by ascending $c_{i,j}$ values. Their new vehicle destination will be the first $i$ parking lot that offers at least one free parking space at the moment of the participating vehicles' departure. This behavior models that commuter drivers usually check parking availability when they depart, or it is similar to the parking guidance systems showing the number of free parking spaces in P+R parkings on traffic signs at the perimeter of the cities.

As this system still does not offer parking reservations, there is a certain risk that a previously free parking space will be occupied while a vehicle is driving towards it. If a driver finds its target parking lot full, it starts cruising as described in section~\ref{sec:cruising_sim}. As we simulate curbside parking, we assume that it is rational that due to traffic, drivers cannot stop safely and recheck the actual parking occupancies after their departure. Naturally, in corner cases, i.e., when a favorable parking lot has only the last spaces free, multiple vehicles might set it as their destination. However, only a small fraction of them can indeed occupy the last spaces. The rest of the vehicles will have to cruise to find a free parking spot, possibly causing a minor traffic congestion.

\subsubsection{Auctions}

\begin{figure}[tb]
    \centering
    \resizebox*{12.5cm}{!}{\includegraphics{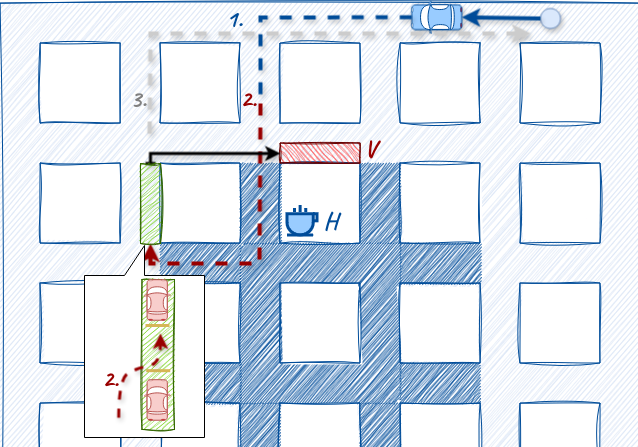}}
    \caption{An example for the simulated vehicle behavior. The vehicle appeared in the simulation at the point marked by a circle. Its driver's human destination is a café right in the center (H). As it lays in the more expensive (darker colored) parking zone, the driver (having a higher $\beta$ preference), opt to the closest parking lot in the cheaper zone (V). In the baseline and in the auction system, the driver starts moving toward V (1.), while by using the information system, it can target to go to another similar space if V has no free parking spaces (6.6~m long free space in a particular parking lot). In the auction system, with a 15~s period time, the (automated bidding agent) of the driver can participate and reserve a parking space, not necessarily in V, and possibly adding to the route length (2.). After parking, the vehicles leave the simulation at their street of origin(3.). The parking distance between the occupied and the preferred parking is depicted by a black arrow.} \label{fig:grid_distance}
\end{figure}

We also implemented the auction-based parking reservation system described in section~\ref{sec:auction}. We scheduled the auctions to run before every 15th simulation step, corresponding to a 15~s period. Consequently, auctions run in each $(15 k)$th step, $k \in \{0, 1, 2, 3, \dots\}$. Participants departing between the $\big(15 (k-1)\big)$th and the $(15 k - 1)$th step shall participate in the auctions starting before the $(15 k)$th simulation execution step. The auction method assigns vehicles to parking lots. To occupy them, we reroute vehicles to parking lots using a Python script via an interface called Traffic Control Interface, \texttt{TraCI}. Due to the 15~s period time of auction running, and accordingly controlling the simulation via \texttt{TraCI}, a handful of vehicles arrive at their original destination prior to this interaction. Hence, there can always be some non-participating vehicles. According to our results, only a negligible 1.5\% of the vehicles have a route shorter than 15~simulation steps in the simulated scenario.

Consequently, there is always a possibility that a reserved parking space gets physically occupied by non-participating vehicles. In this case, the participating vehicles will also start cruising to find a vacant parking space by the traditional method, see section~\ref{sec:cruising_sim}.

\subsection{Simulation Execution}
\label{sec:sim_ex}

We ran simulations with a 1~s long discrete step time. 
We tested the system with 6~different penetration rates for both the information and auctions parking behavior, ranging from 0\% (baseline) to 100\% with steps of 20\%. 

We repeated each simulation case 10~times, in which vehicles departed with a (uniformly) random departure offset of 5~minutes with fixed random seeds. This random seeds primarily affect the departure time of the vehicles; hence, creating slightly different simulation scenarios with different interactions between the vehicles and the infrastructure. Consequently, we can ensure that the results are statistically stable and reproducible.

\subsection{Measured Parameters}
To evaluate the capabilities of the proposed system, we measured the following values in the Eclipse SUMO simulations.

\emph{Parking prices.} When the vehicles stopped for parking in the simulations, we recorded the parking price they had to pay depending on the parking behavior. For the baseline and the information system, the resulting parking prices correspond to the parking prices in the parking zones. In the case of auction parking behavior, the parking price corresponds to the winning price on the auction, provided that drivers successfully occupy their reserved parking space. Otherwise, the paid parking price, similarly to the baseline and information case, is defined by the parking zones.

Moreover, in the case of the auction system, participating and non-participating vehicles are charged by different parking pricing schemes. To analyze this difference, we also separately checked the parking prices for both groups. Additionally, we provide statistics on the rate of successful reservations at different levels of market penetration for the auction system. We expect that the auction system will lower the price difference at the edges of the parking zones.

For individuals, lower parking prices are preferable; however, parking operators would not be satisfied with the novel parking system if their revenues were to decrease. 

\emph{Parking distance.} With the help of \texttt{TraCI} in Eclipse SUMO, when a vehicle stops for parking, we check the distance between the actually occupied parking space and the driver's original vehicular destination (i.e., its most preferred place for parking), see Figure~\ref{fig:grid_distance}. This distance represents the driving distance from the occupied parking space to the most preferred one, which is an overestimate for walking distance, as there may be shortcuts or pathways between these two points. Unfortunately, Eclipse SUMO cannot simulate reversing or turning around. This functionality can naturally increase the driving distance further. A lower parking distance is better.

\emph{Route length.} When vehicles leave the simulation, we collect their total route length. Cruising for parking can significantly increase the total route length; hence, it can be a qualitative indicator of the efficiency of the parking systems. A relatively shorter route length is better.

\emph{Parking occupancy patterns.} After the simulations complete, we aggregate parking lot occupancies over time to obtain the average level of parking occupancy for each edge of the road network. A prominent change in the occupancy patterns indicates that the introduced parking system changes drivers' parking habits, and possibly causes inconvenience as \emph{human} destinations are unchanged, and still focused on the center of the CBD. Moreover, the parking information system and the auction-based parking reservation system can balance parking lot occupancies, which is beneficial because it mitigates the overload of certain parking lots, allowing a vehicle to find a vacant space on the preferred street.

\emph{Traffic flow.} In addition to parking simulation, we can also measure traffic flow in Eclipse SUMO. To do this, we placed induction loop traffic detectors\footnote{One can find additional details in Eclipse SUMO documentation:\\ \url{https://sumo.dlr.de/docs/Simulation/Output/Induction_Loops_Detectors_(E1).html}} on each road edge of the simulated network, 40~m apart from its downstream end. These detectors provide traffic flow data with a 900~s period in vehicles/hour ([veh/h]) unit. For a compact view of the average traffic flow in the road network, we aggregate the traffic flows over the simulation time and across all edges in the road network. As each simulation case has a burn-in and run-down phase, we cut and analyze only the steady-state section of the traffic flow patterns. A higher traffic flow indicates a higher network throughput, hence a more efficient transportation system.

\section{Results with Different Penetration Rates}
\label{sec:results}
\begin{figure}[tb]
    \centering
    \resizebox*{14cm}{!}{\includegraphics{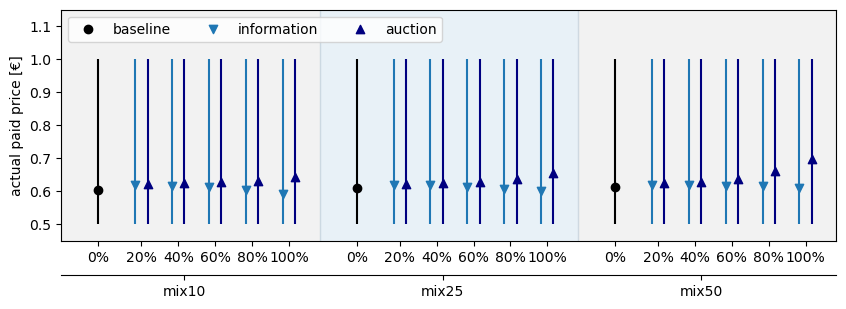}}
    \caption{Parking prices in various measurement cases, parking behavior and market penetration levels. Markers show the mean values, whiskers range between the 10th and 90th percentiles.} \label{fig:prices}
\end{figure}

\begin{figure}[tb]
    \centering
    \resizebox*{14cm}{!}{\includegraphics{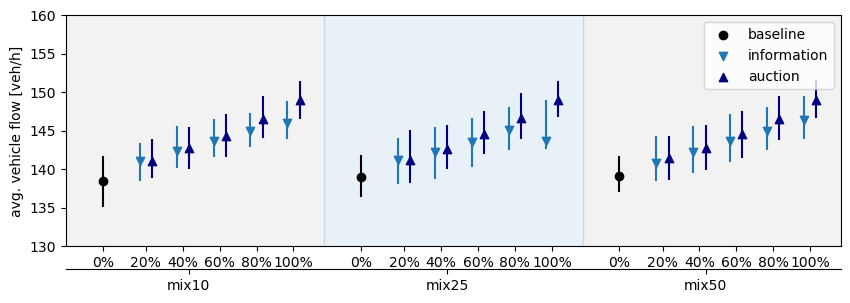}}
    \caption{Average traffic flow in various measurement cases, parking behavior and market penetration levels. Markers show the mean values, whiskers range between the 10th and 90th percentiles.} \label{fig:flows}
\end{figure}

The simulation results allow us to analyze how auction-based parking assignment affects the traffic system and the parking economy at different levels of market penetration. In general, we can observe that both an information system and the auction-based system can improve traffic flow, see Table~\ref{tab:results} and Figure~\ref{fig:flows}. Moreover, the prices tend to increase with the auction-based system and decrease with the information system, see Table~\ref{tab:results} and Figure~\ref{fig:prices}.

In the following, we check how different rate of penetration affects the measured parameters.

\begin{table}[tb]
    \tbl{Mean numerical results for various measurement cases and parking behaviors at different levels of market penetration.}
    {\begin{tabular}{l|l||r|rrrrr|rrrrr} \toprule
     && \textbf{baseline} & \multicolumn{5}{c}{\textbf{information}} & \multicolumn{5}{c}{\textbf{auction}} \\ 
     && & 20\% & 40\% & 60\% & 80\% & 100\% & 20\% & 40\% & 60\% & 80\% & 100\% \\ \midrule
     \multirow{4}{*}{\rotatebox{90}{\texttt{MIX10}}} & route length [m] & 553 & 564 & 570 & 574 & 579 & 585 & 564 & 573 & 577 & 579 & 583 \\
     & parking price [\euro]& 0.604 & 0.620 & 0.617 & 0.611 & 0.603 & 0.592 & 0.623 & 0.624 & 0.627 & 0.632 & 0.641 \\ 
     & parking distance [m] & 125 & 116 & 111 & 110 & 115 & 128 & 115 & 103 & 88 & 70 & 49 \\
     & flow [veh/h] & 138.4 & 141.0 & 142.4 & 143.6 & 145.0 & 146.0 & 141.1 & 142.7 & 144.3 & 146.5 & 148.9 \\ \midrule
     \multirow{4}{*}{\rotatebox{90}{\texttt{MIX25}}} & route length [m] & 557 & 565 & 569 & 574 & 579 & 583 & 565 & 570 & 578 & 585 & 594 \\ 
     & parking price [\euro] & 0.608 & 0.620 & 0.618 & 0.614 & 0.608 & 0.599 & 0.623 & 0.624 & 0.628 & 0.638 & 0.654 \\
     & parking distance [m] & 115 & 114 & 109 & 106 & 106 & 109 & 114 & 102 & 88 & 70 & 50 \\
     & flow [veh/h] & 139.0 & 141.2 & 142.2 & 143.5 & 145.1 & 143.6 & 141.2 & 142.6 & 144.6 & 146.7 & 148.9 \\ \midrule
     \multirow{4}{*}{\rotatebox{90}{\texttt{MIX50}}} & route length [m] & 557 & 564 & 569 & 574 & 579 & 584 & 565 & 570 & 577 & 584 & 593 \\
     & parking price [\euro] & 0.614 & 0.620 & 0.619 & 0.617 & 0.614 & 0.611 & 0.624 & 0.628 & 0.637 & 0.660 & 0.697 \\
     & parking distance [m] & 115 & 118 & 112 & 106 & 100 & 97 & 114 & 103 & 87 & 71 & 51 \\ 
     & flow [veh/h] & 139.1 & 140.9 & 142.2 & 143.7 & 144.9 & 146.3 & 141.4 & 142.7 & 144.5 & 146.6 & 149.0 \\ \bottomrule
    \end{tabular}}
    \label{tab:results}
\end{table}

\begin{table}[tb]
    \tbl{Mean numerical results for various measurement cases for participating and not participating drivers at different levels of market penetration of the auction-based parking assignment system.}
    {\begin{tabular}{l|l||rrrrr|rrrrr} \toprule
     && \multicolumn{5}{c}{\textbf{non-participants}} & \multicolumn{5}{c}{\textbf{participants}} \\ 
     && 0\% & 20\% & 40\% & 60\% & 80\% & 20\% & 40\% & 60\% & 80\% & 100\% \\ \midrule
     \multirow{3}{*}{\rotatebox{90}{\texttt{MIX10}}} & route length [m] & 553 & 558 & 558 & 557 & 559 & 585 & 589 & 590 & 591 & 594 \\
     & parking price [\euro]& 0.604 & 0.626 & 0.631 & 0.638 & 0.644 & 0.609 & 0.612 & 0.619 &  0.629 & 0.641\\ 
     & parking distance [m] & 124 & 123 & 119 & 112 & 105 & 86 & 80 &72 & 61 & 49 \\
     \midrule
     \multirow{3}{*}{\rotatebox{90}{\texttt{MIX25}}} & route length [m] & 557 & 559 & 559 & 558 & 560 & 587 & 5588 & 590 & 592 & 594 \\ 
     & parking price [\euro] & 0.608 & 0.626 & 0.630 & 0.636 & 0.641 & 0.610 & 0.616 & 0.629 & 0.637 & 0.654 \\
     & parking distance [m] & 118 & 121 & 118 & 112 & 102 & 83 & 79 & 71 & 62 & 50 \\
 \midrule
     \multirow{3}{*}{\rotatebox{90}{\texttt{MIX50}}} & route length [m] & 557 & 559 & 558 & 559 & 557 & 586 & 589 & 590 & 591 & 593 \\
     & parking price [\euro] & 0.614 & 0.626 & 0.630 & 0.635 & 0.640 & 0.613 & 0.624 & 0.638 & 0.665 & 0.679\\
     & parking distance [m] & 114 & 122 & 119 & 111 & 104 & 85 & 80 & 71 & 62 & 50 \\  \bottomrule
    \end{tabular}}
    \label{tab:groups}
\end{table}

\begin{figure}[tb]
    \centering
    \resizebox*{9cm}{!}{\includegraphics{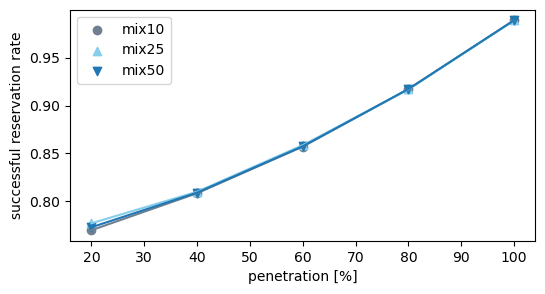}}
    \caption{Rate of successful physical parking reservation with the auction-based system on different market penetration levels.} \label{fig:succ_res}
\end{figure}

\begin{figure}[tb]
    \centering
    \resizebox*{14cm}{!}{\includegraphics{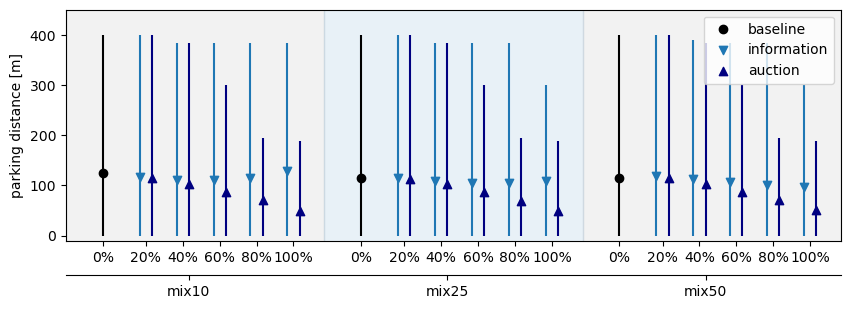}}
    \caption{Parking distances in various measurement cases, parking behavior and market penetration levels. Markers show the mean values, whiskers range between the 10th and 90th percentiles.} \label{fig:distance}
\end{figure}

\begin{figure}[tb]
    \centering
    \resizebox*{14cm}{!}{\includegraphics{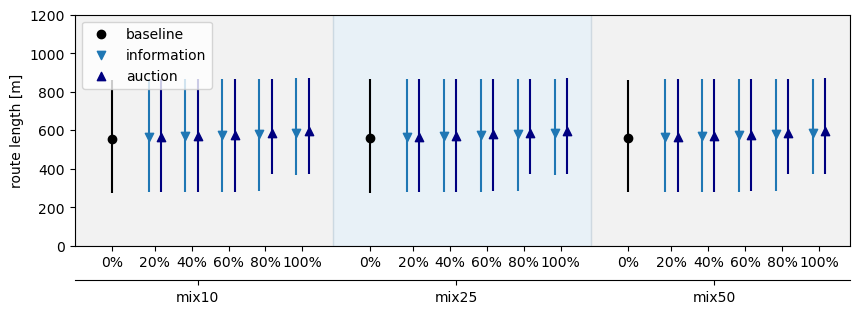}}
    \caption{Route lengths in various measurement cases, parking behavior and market penetration levels. Markers show the mean values, whiskers range between the 10th and 90th percentiles.} \label{fig:lengths}
\end{figure}

\begin{figure}[tb]\centering
    \begin{subfigure}{\linewidth}
        \centering
        \resizebox*{14cm}{!}{\includegraphics{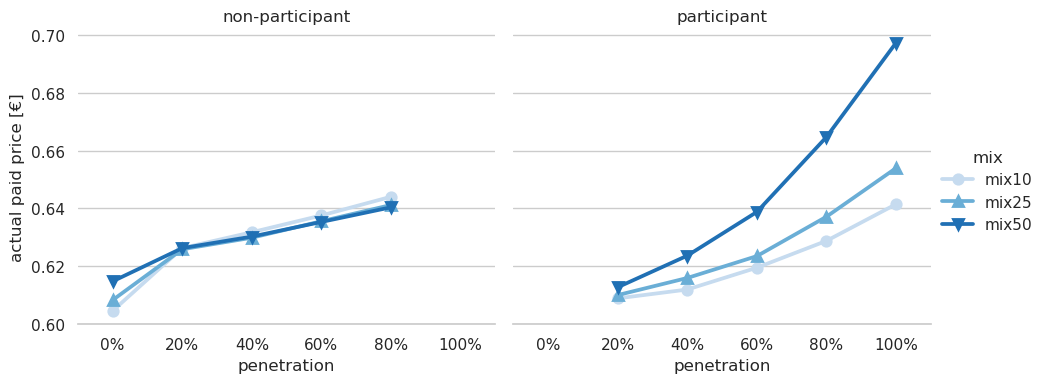}}
        \caption{Parking prices.}
        \label{fig:auc_prices}
    \end{subfigure}
    
    \begin{subfigure}{\linewidth}
        \centering
        \resizebox*{14cm}{!}{\includegraphics{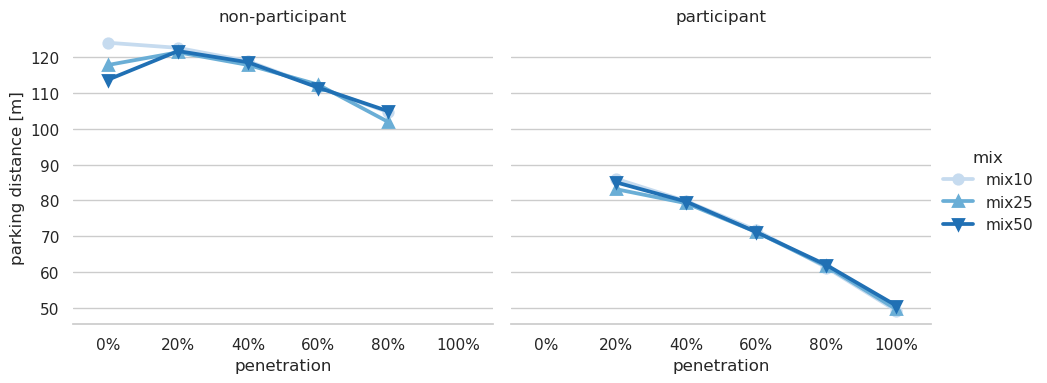}}
        \caption{Parking distances.}
        \label{fig:parking_distance_pen}
    \end{subfigure}
    
    \begin{subfigure}{\linewidth}
        \centering
        \resizebox*{14cm}{!}{\includegraphics{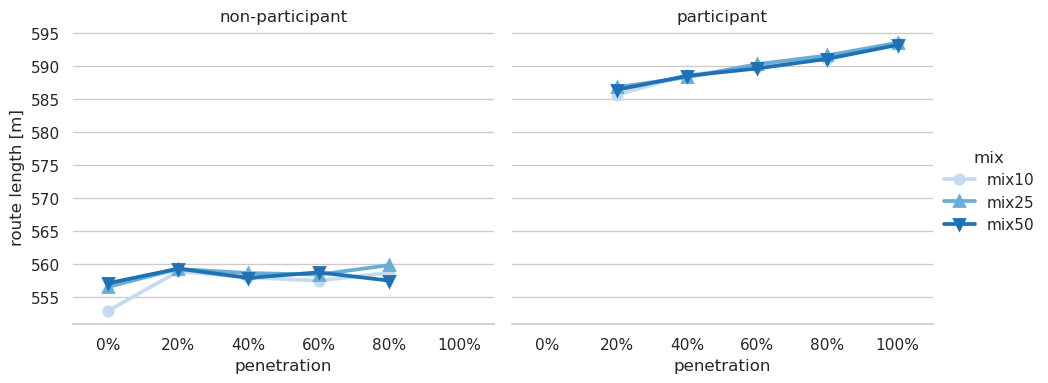}}
        \caption{Route lengths.}
        \label{fig:route_length_pen}
    \end{subfigure}
    
    \caption{Various results of the auction-based reservation system grouped by participating and non-participating vehicles in various measurement cases, with the auction-based parking reservation method on different market penetration levels. Markers show the mean values.}
    \label{fig:grouped}
\end{figure}

\begin{figure}[tb]
    \centering
    \resizebox*{14cm}{!}{\includegraphics{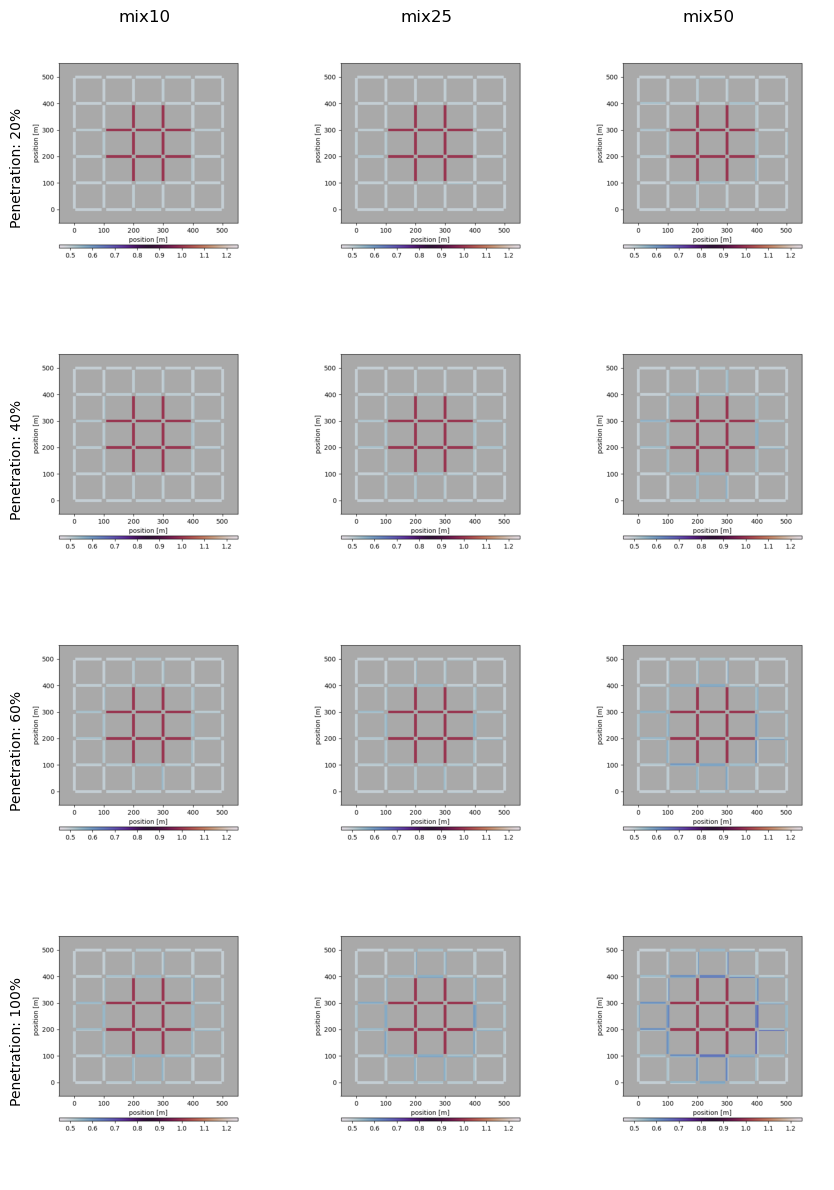}}
    \caption{Parking price patterns of the auction parking behavior on different penetration rates. Lighter blueish colors corresponds to lower, red colors to higher parking prices.} \label{fig:auc_price}
\end{figure}

\begin{figure}[tb]
    \centering
    \resizebox*{14cm}{!}{\includegraphics{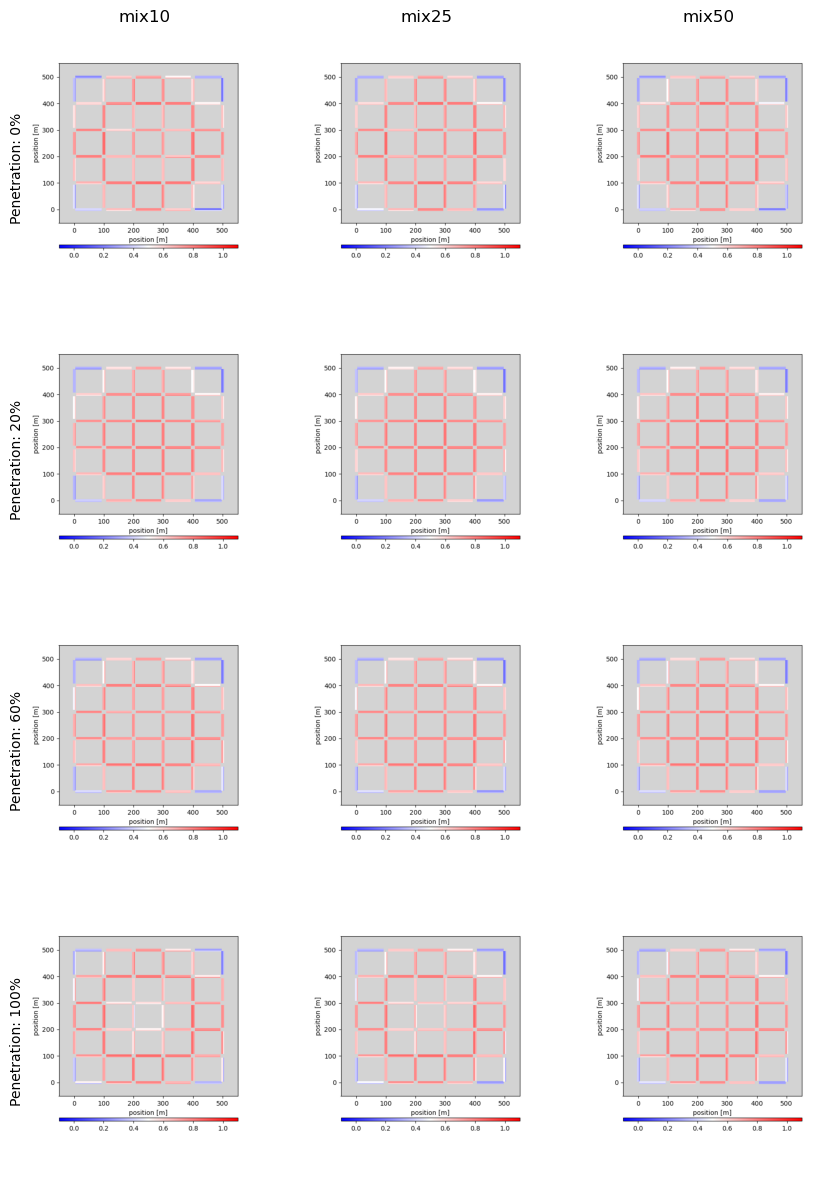}}
    \caption{Occupancy patterns of the information system parking behavior on different penetration rates. Red colors corresponds to higher, blue colors to lower occupation rates.} \label{fig:info_occup}
\end{figure}

\begin{figure}[tb]
    \centering
    \resizebox*{14cm}{!}{\includegraphics{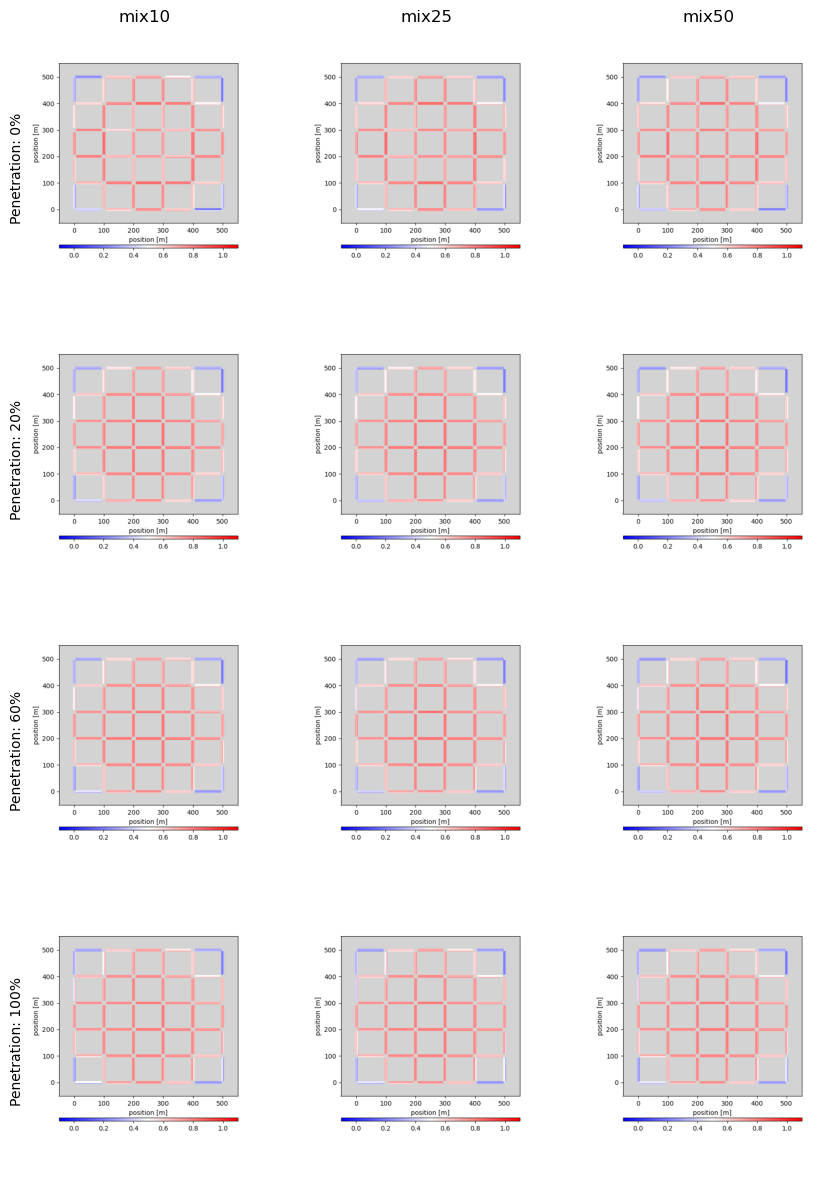}}
    \caption{Occupancy patterns of the auction parking behavior on different penetration rates. Red colors corresponds to higher, blue colors to lower occupation rates.} \label{fig:auc_occup}
\end{figure}

\subsection{Lower Penetration Levels}

At lower penetration levels ($\leq 40$\%), vehicles experience an insignificant increase in route length in each measurement case; see Figure~\ref{fig:lengths}. and Table~\ref{tab:results}. In the meantime, they can get slightly closer to their preferred parking place compared to the baseline scenario regardless of the measurement case, according to Figure~\ref{fig:distance}. and Table~\ref{tab:results}. We can also observe that the improvement is slightly higher with the auction-based reservation system.

As shown in Figure~\ref{fig:prices} and Table~\ref{tab:results}, parking prices increase with both the information and auction systems compared to baseline behavior. The auction system also reflects the differences between the measurement cases; the increment gets higher as the number of drivers with lower $\beta$s (price aversion) gets lower. According to Figure~\ref{fig:auc_price}., there is no visible change in parking prices in different parking lots at low penetration levels. However, at low penetration levels, participating drivers will pay generally less parking fees compared to non-participating drivers, regardless of the measurement case, see Figure~\ref{fig:auc_prices}. and Table~\ref{tab:groups}. As the rate of successful reservations is around 80\% even at low penetration rates, see Figure~\ref{fig:succ_res}., this phenomenon is most likely caused by the fact that the non-participating drivers cannot occupy their desired parking space (likely on the edge of the parking zones). Instead, non-participating drivers shall park in the more expensive zone, while participating vehicles can reserve a cheaper alternative for themselves.

However, participating drivers can reduce the parking distance significantly with the auction system, even at low penetration levels, the route length increases, see Figure~\ref{fig:parking_distance_pen}., Figure~\ref{fig:route_length_pen}., and Table~\ref{tab:groups}. The route length increment is due to the possible rerouting to the reserved parking lot, see Figure~\ref{fig:grid_distance}.

At low penetration levels, both the information and auction systems provide a more balanced parking occupancy pattern across the road network compared to the baseline (0\% penetration case), see Figure~\ref{fig:info_occup}., and Figure~\ref{fig:auc_occup}.

It is also a surprising effect that even at low penetration levels, the traffic flow  improves in each measurement case of both the information and auction-based reservation systems, see Figure~\ref{fig:flows}. and Table~\ref{tab:results}.

\subsection{Medium Penetration Levels}
At medium penetration levels (40--80\%), previous trends seem to continue. Parking distance significantly decreases with auction system in every measurement case, while the information system results in smaller improvements, see Table~\ref{tab:results} and Figure~\ref{fig:distance}. According to Figure~\ref{fig:succ_res}, this is due to the approximately 90\% success rate of the reservation system.

With the auction system, the parking prices tend to equalize at the edge of the parking zones, see Figure~\ref{fig:auc_price}., and we can see a significant increase in auctioned parking prices in Figure~\ref{fig:auc_prices}. However, depending on the measurement case; participants can still find cheaper parking lots than non-participants, see Table~\ref{tab:groups}. Interestingly, the parking assignment system also helps reducing non-participant drivers' parking distance at medium penetration levels, see Figure~\ref{fig:parking_distance_pen}. Considering the on 40\% of penetration, the parking distances are similar for non-participants to the baseline case, and on 60\%, it shows some improvement; we can suspect that if at least 50\% of the drivers use the reservation system, even the non-participants can be more successful in occupying their preferred parking spots.

At the medium penetration levels, both the auction and the information system can balance parking lot occupancies in the road network compared to the baseline behavior, see Figure~\ref{fig:info_occup}. and Figure~\ref{fig:auc_occup}. Moreover, both systems can significantly improve the traffic flow, according to Figure~\ref{fig:flows}. and Table~\ref{tab:results}.

\subsection{High penetration levels}
At complete penetration (80-100\%), the information system's performance is likely to degrade compared to lower rate of penetration. It is due to the effect mentioned in section~\ref{sec:info_behav}, that the information system indicates free parking spaces at specific points at the moment of the departure of the vehicles, but they get occupied while the vehicles are driving towards them. This creates minor congestion, especially in case of the \texttt{MIX25} scenario, see the decrease in traffic flow in Table~\ref{tab:results}, and in Figure~\ref{fig:flows}. We shall note, that the traffic flow is still higher than it used to be with the baseline behavior.

Moreover, at complete penetration level, the information system almost reproduces the original (baseline) parking occupancy patterns across the road network, see Figure~\ref{fig:info_occup}. Together with the decreased traffic flow in the \texttt{MIX25} case, it could be an indication that the information system might struggle to fulfill its task in certain situations.

Regarding the auction-based reservation system, it exceeds 95\% of the successful reservation rate, see Figure~\ref{fig:succ_res}, but due to the 15~s period time of the auction runs, described in section~\ref{sec:sim_ex}, it cannot reach 100\%.\footnote{In our simulations, it achieves an average of 98.5\%.} Consequently, the parking prices also increase by 7--14\%, depending on the measurement case, see Figure~\ref{fig:prices}. and Table~\ref{tab:results}. This price increment, especially in the case of \texttt{MIX50}, also results in a less steep price gradient at the border of the original two parking zones, see Figure~\ref{fig:auc_price}, while also balances the parking occupancy pattern, see Figure~\ref{fig:auc_occup}.

According to Table~\ref{tab:results}., the auction-based parking reservation system with 100\% penetration improves traffic flow the most in each measurement case.

Additionally, the complete penetration of the auction system can halve the parking distance even when it is compared the best results of the information system in the \texttt{MIX10} and \texttt{MIX25} cases, see Table~\ref{tab:results}. and Figure~\ref{fig:distance}, and reduces parking distances by 55--60\% compared to the original baseline behavior depending on the measurement case.

An interesting result is that the route lengths are not shortened by either the information nor the auction system. Due to the already described feature of information system, it can add cruising when it indicates free parking lots, but they get occupied before some of the vehicles could arrive. On the other hand, for the auction-based system its increment is caused when we influence the simulation after the auctions complete. As we reroute the vehicles to their assigned parking lots, they might have to make a detour to reach these streets, which results in an increased route length. However, by implementational improvements, i.e., by reducing the rate of the auction period time over the vehicles' travel time, this route length increment could be decreased.

\section{Conclusions}
\label{sec:conclusion}
This paper focused on filling a research gap by exploring various aspects of the transition from a traditional to an intelligent, auction-based parking assignment system. Assuming that the developers deploy the proposed system as a mobile app, the drivers can decide whether or not they would like to use it. As the rate of the participant and non-participant users can affect both the traffic and the performance of the assignment system, we tested the assumed system at various market penetration levels with an Eclipse SUMO simulation.

The novel system can significantly increase the flow of traffic with increasing penetration levels. There is even a visible improvement at low penetration levels, which could mitigate traffic congestion when only a small fraction of drivers participate in the novel system. Auction-based parking reservation, in contrast to a simple parking information system, can provide constant improvements as its penetration rate grows.

We assumed that non-participant drivers could occupy some spaces already assigned to participants. However, even on a low penetration level, participants can occupy their designated parking lots in most cases, according to the results. In addition to having a reserved parking lot, at lower penetration levels, participant drivers can pay lower parking fees compared to the non-participants. These benefits will likely attract more users, further increasing penetration.

For an auction-based assignment, we might expect that the actual paid parking prices will increase. However, participants' parking expenditure does not only exceed the traditional costs by 7--14\% in our various measurement cases. On the other hand, with the auction system's increasing parking prices, in contrast to the decrease with a simple information system, parking lot operators would be also motivated to further increase the auction system's market penetration to also increase their revenues.

Together, an auction-based parking assignment system solves the century-old problem of market-driven parking pricing. Moreover, it also provides many benefits to its stakeholders. We conclude that the system is operational even with low penetration rates; therefore, it might be worth it to implement and deploy such a parking assignment system. According to the simulation results, the system can potentially gain a significant market penetration rate. However, before deployment, in addition to developers, parking operators, i.e., municipalities, must also make simulations and evaluate the results in their specific case, including the road network of a city, a calibrated traffic and parking model, and original real-world parking fees. Furthermore, one can also implement different auction preference functions similar to our current research plan. To encourage these experiments and tests, we have published our project on GitHub: 
\url{https://github.com/alelevente/penparking}.

\section*{Funding}
This research was supported by project no. EKÖP-24-3-BME-319, implemented with the support provided by the Ministry of Culture and Innovation of Hungary from the National Research, Development and Innovation Fund, financed under the EKÖP-24-3 funding scheme; as well as by project no. TKP2021-EGA-02, implemented with the support provided by the Ministry of Culture and Innovation of Hungary from the National Research, Development and Innovation Fund, financed under the TKP2021-EGA funding scheme.

\bibliographystyle{apacite}
\bibliography{./interactapasample.bib}

\end{document}